\documentclass[conference]{IEEEtran}
\IEEEoverridecommandlockouts
\usepackage{cite}
\usepackage{amsmath,amssymb,amsfonts}
\usepackage{algorithmic}
\usepackage{graphicx}
\usepackage{textcomp}
\usepackage{xcolor}
\usepackage{booktabs} 
\usepackage{acronym} 
\usepackage{hyperref}
\usepackage{flushend} 
\usepackage{tikz}

\newcommand\copyrighttext{%
  \footnotesize \textcopyright 2025 IEEE. Personal use of this material is permitted.
  Permission from IEEE must be obtained for all other uses, in any current or future
  media, including reprinting/republishing this material for advertising or promotional
  purposes, creating new collective works, for resale or redistribution to servers or
  lists, or reuse of any copyrighted component of this work in other works.}
\newcommand\copyrightnotice{%
\begin{tikzpicture}[remember picture,overlay]%
\node[anchor=south,yshift=10pt] at (current page.south)%
  {\fbox{\parbox{\dimexpr\textwidth-\fboxsep-\fboxrule\relax}{\copyrighttext}}};
\end{tikzpicture}%
}
\acrodef{BD}{Backdoor Deployment}
\acrodef{EC}{Escalation of Control}
\acrodef{ER}{Exposure and Removal}
\acrodef{ML}{machine learning}
\acrodef{LC}{Legitimate Contributions}
\acrodef{SBOM}{Software Bill of Materials}
\acrodef{SSC}{software supply chain}
\acrodef{MAPE-K}{Monitor, Analyse, Plan, Execute, Knowledge}

\begin{document}
\title{Towards Socio-Technical Topology-Aware Adaptive Threat Detection in Software Supply Chains}

\author{
\IEEEauthorblockN{Thomas Welsh, Kristófer Finnsson, Brynjólfur Stefánsson and Helmut Neukirchen}
\IEEEauthorblockA{Department of Computer Science, University of Iceland, Reykjavík, Iceland\\
\{tomwelsh, kdf2, brs, helmut\}@hi.is}
}

\maketitle
\copyrightnotice
\begin{abstract}
\Acp{SSC} are complex systems composed of dynamic, heterogeneous technical and social components which collectively achieve the production and maintenance of software artefacts. Attacks on \acp{SSC} are increasing, yet pervasive vulnerability analysis is challenging due to their complexity. Therefore, threat detection must be targeted, to account for the large and dynamic structure, and adaptive, to account for its change and diversity. While current work focuses on technical approaches for monitoring supply chain dependencies and establishing component controls, approaches which inform threat detection through understanding the socio-technical dynamics are lacking. We outline a position and research vision to develop and investigate the use of socio-technical models to support adaptive threat detection of \acp{SSC}. We motivate this approach through an analysis of the XZ Utils attack whereby malicious actors undermined the maintainers' trust via the project's GitHub and mailing lists. We highlight that monitoring technical and social data can  identify trends which indicate suspicious behaviour to then inform targeted and intensive vulnerability assessment. We identify challenges and research directions to achieve this vision considering techniques for developer and software analysis, decentralised adaptation and the need for a test bed for software supply chain security research.
\end{abstract}


\acresetall 

\section{Introduction}
The \ac{SSC} involves a complex network of interdependent components from initial requirements gathering, to deployment upon an end-user system. Modern software development heavily employs code reuse to enhance productivity, this results in code making use of a network of dependencies, most of which are opaque to the developer and user. This interdependency ensures attacks against one component may propagate downstream \cite{okafor2022sok} with \ac{SSC} attacks,  such as malicious packages, increasing by 156\% yearly \cite{Sonatype2024}. These components are under continuous evolution and independently maintained by different organisations and individuals operating within various geographical locations and legislative environments \cite{okafor2022sok}. This requires threat detection and security controls to evolve alongside them and to perform continuous evaluation within varying social and legislative environments.
\ac{SSC} attacks have been highlighted as a key challenge by the security and software engineering communities \cite{enck2022top} due to the increasing number of high-profile incidents \cite{alkhadra2021solar,feng2022defense}, their impact upon a wide number of end-users, and the challenges in detecting attacks relating to the complexity of the \ac{SSC}.


Typically, approaches to secure the \ac{SSC} focus upon dependency monitoring, enhancing secure development practices, or applying controls to each supply chain stage. These approaches do not aim to mitigate social attack vectors and their effectiveness in mitigating them is unknown \cite{williams2025research, enck2022top}. 

In contrast, this paper proposes an alternative approach which leverages the complexity characteristics of the \ac{SSC}, as opposed to building controls to mitigate them. Through monitoring and adapting to changes in: social behaviours, components and structural properties, we suggest that threat detection can operate in a targeted  manner to detect complex and evolving social threats. We propose to generate and monitor \textit{socio-technical topologies of the \ac{SSC}} to support analysis of socio-technical threat indicators. To manage change, these topologies can be analysed within an adaptive framework, resulting in \textit{Socio-Technical Adaptive \ac{SSC} Threat Detection}. Expanding on a previous approach for cyber-physical supply chains \cite{welsh2022topology}, yet applying techniques for modelling and analysing socio-technical dynamics specific to software development.

Towards achieving this vision, the contributions of this position paper are as follows: 1) We motivate the need for socio-technical topology modelling and analysis to support threat detection of \acp{SSC} through a study of the XZ Utils attack \cite{przymus2025wolves}. We highlight that aggregating social and technical data sources could indicate suspicious developer behaviour. 2) We propose a framework to adaptively detect threats in \acp{SSC}  supported by socio-technical topologies. 

The rest of this paper is structured as follows. Section~\ref{sec:Background and Related Work} provides a background in \ac{SSC} security, socio-technical security, and socio-technical modelling. A motivating example of the XZ Utils \ac{SSC} attack is presented in Section~\ref{sec:Motivating Example: The XZ Utils Supply Chain Attack}. Section~\ref{sec:Socio-Technical Adaptive Threat Detection} presents the proposed approach, \textit{Socio-Technical Adaptive SSC Threat Detection}. Section~\ref{sec:Limitations and Research Vision} provides a discussion, followed by a conclusion in Section~\ref{sec:Conclusion}.

\section{Background and Related Work}\label{sec:Background and Related Work}
Dependency-focused attacks vectors are common to \ac{SSC} security \cite{enck2022top}. The \ac{SBOM} is a key mitigator but suffers from numerous challenges  \cite{enck2022top, stalnaker2024boms}. Version control systems, build systems, package distribution systems, and end-user machines are also targets of attack \cite{ladisa2023sok}. Threats from social factors are also numerous, e.g.\ malicious insiders injecting code \cite{lins2024critical} or from a variety of psychological phenomena \cite{iannone2022secret, rauf2022influences, ivory2023recognizing}. 
Consequently, software repositories can be mined for social indicators of code quality,~\cite{duenas2018perceval, german2016continuously}, bug reports can predict vulnerabilities \cite{peters2017text, lopez2019anatomy}, version control systems and mailing lists can link developers with  code \cite{carlson2015engaging} and chats can indicate project sentiment \cite{kuutila2020chat}. Where anomalous code deviations, sentiment or bug reports indicate a vulnerable software component. \Ac{ML} approaches show success in monitoring individual components but not the complete supply chain or its change. They may also suffer from high rates of false positives \cite{10.1145/3092566}. Quantum \ac{ML} has shown to offer lower accuracy and higher computation than classical approaches \cite{masum2022quantum}.
The wide social and technical attack surface suggests that \acp{SSC} should be considered as a socio-technical system \cite{al2015socio, baxter2011socio}. Threats arising from vulnerabilities in human, social and organisational factors are widely known, e.g.\ social engineering \cite{syafitri2022social}, inadequate organisational incident response preparation \cite{caldwell2012prepare}, and ineffective security culture \cite{gcaza2017cybersecurity}. Socio-technical security considers that unique vulnerabilities arise due to the interplay of both social and technical vulnerabilities \cite{goerzen2019entanglements}. To identify unique socio-technical threats, software-based threat detection must have the means to model and reason about these concepts cohesively.

Socio-technical systems are complex and therefore challenging to model, monitor, and analyse due to their emergent, heterogeneous, dynamic and large structures \cite{vespignani2012modelling}. Network-theoretic approaches are frequently used to understand these structures \cite{estrada2012structure}. Topology modelling  is used to combine multiple dimensions, (e.g.\ cyber, physical, social) to allow cross-dimensional analysis of structural properties \cite{welsh2022topology,toth2021inequality,alrimawi2019automated}. However, there is a current lack of techniques which investigate the use of socio-technical topologies of \acp{SSC}.

Pervasive monitoring and inspection of all components across a dynamically evolving \ac{SSC} is cost ineffective, creating a resource allocation problem \cite{welsh2022topology}. While topologies can be used to manage the structural complexity of a complex system, further techniques are needed to handle this change and emergent behaviours. Adaptive software is used to manage change, e.g.\ based on the \ac{MAPE-K} adaptive software framework \cite{quin2021decentralized}, and was previously used for threat detection in cyber-physical supply chains \cite{welsh2022topology}. We propose this approach can also apply to threat detection in \acp{SSC} yet including novel techniques for modelling socio-technical factors of developers 
to inform targeted analysis of software components for vulnerabilities.

\section{Motivating Example: The XZ Utils Supply Chain Attack And Threat Indicators}\label{sec:Motivating Example: The XZ Utils Supply Chain Attack}

In this section, we present an initial analysis of the social and technical factors related to the XZ Utils attack \cite{Thompson2024, Kaspersky2024_2}.

\subsection{XZ Utils Attack Stages}\label{subsec:XZ Utils Attack Stages}

The XZ Utils attack can be categorised into several stages:

\noindent
\textbf{\ac{LC}} \textbf{Late '21 – Early '22:} Author \textit{Jia Tan} begins contributing small patches through the xz-devel mailing list, focusing on minor bug fixes and feature improvements. In total, \textit{Jia Tan} authored over 500 patches in multiple GitHub projects. These early contributions appear legitimate, gradually establishing credibility.

\noindent
\textbf{\ac{EC}} \textbf{Early '22 – Early '24:} Lead maintainer \textit{Lasse Collin} is placed under sustained social pressure via mailing list discussions. Multiple anonymous or sockpuppet accounts criticize slow project progress, gradually pressuring him into relinquishing control. \textbf{Mid '22 – \mbox{Early '23:}} \textit{Jia Tan} is recognized in Git metadata as an author, eventually becoming a co-maintainer with commit access. \textbf{March '23:} \textit{Jia Tan} tags and builds the first release under own control (\textit{v5.4.2}), marking a critical shift in project leadership. \textbf{June '23 – January '24:} Technical groundwork is laid for the attack: \textbf{June '23:} \textit{Hans Jansen}, a new actor, submits patches introducing the GNU indirect function (\textit{ifunc}) feature, which is later leveraged for the backdoor.
\textbf{July '23:} \textit{Jia Tan} disables ifunc support in OSS-Fuzz builds, reducing external visibility. \textbf{January '24:} Control over the XZ Utils website is transferred to GitHub Pages, under \textit{Jia Tan’s} control.

\noindent
\textbf{\ac{BD}} \textbf{February '24 – March '24:} \textit{Jia Tan} introduces a subtly obfuscated backdoor into the codebase, hidden within binary test input files. The README file discourages detailed scrutiny, making detection difficult.

\noindent
\textbf{\ac{ER}} \textbf{March 28, '24:} Security researcher \textit{Andres Freund} identifies anomalous SSH behaviour and privately notifies \textit{Debian} and \textit{RedHat}. \textbf{March 29, '24:} \textit{RedHat} assigns \textit{CVE-2024-3094} \cite{cve_record}, marking the vulnerability as a critical security risk. \textbf{March 30, '24:} A public backdoor warning is issued, and \textit{Fedora} confirms that compromised XZ Utils versions were included in a recent release.

\subsection{Socio-Technical Threat Indicators}

\begin{table}[b]\vspace*{-4.0ex}
\caption{File-Level Line Change Statistics: Lines added/deleted by commits from Jia Tan vs.\ Lasse Collin from 2022-01 to 2024-06}\vspace*{-1ex}
\label{tab:comparison_stats}
\centering
\begin{tabular}{lccccc}
\toprule
\textbf{Statistic} & \textbf{Jia Tan} & \textbf{Lasse Collin} \\
\midrule
Total File Changes & 697 & 1973 \\
Average Additions & 89.42 & 28.26 \\
Average Deletions & 42.10 & 18.31  \\
Average Total Changes & 131.53 & 46.56 \\
Std Dev Additions & 396.20 & 146.41  \\
Std Dev Deletions & 163.45 & 147.61  \\
Std Dev Total Changes & 492.14 & 249.01 \\
\bottomrule
\end{tabular}
\end{table}

The complex technical backdoor was achieved through social engineering by \emph{Jia Tan} who was seemingly acting as a normal developer [stage \ac{LC}, see subsection~\ref{subsec:XZ Utils Attack Stages}]. This poses the question \textit{What types of developer behaviour can be used to indicate the software is trending towards an insecure state?} To study this, we mined the XZ Utils GitHub commits and issues~\cite{XZGH2025} and mailing list \cite{XZMailingListArchive} from the start of the project in 2008 to 2025. We examined various sources of developer activity, focusing on the threat actor, \emph{Jia Tan}, with a comparison with known legitimate maintainers. Examining commit metrics, Table~\ref{tab:comparison_stats} presents statistics for file-level Git commit statistics between \emph{Jia Tan} and \emph{Lasse Collin} for the period \mbox{2022-01} to \mbox{2024-06}. Despite \emph{Lasse Collin} affecting more files, \emph{Jia Tan} made, on average, more substantial changes to each file than \emph{Lasse Collin}.

\begin{figure}
    \centering
    \includegraphics[scale=0.27]{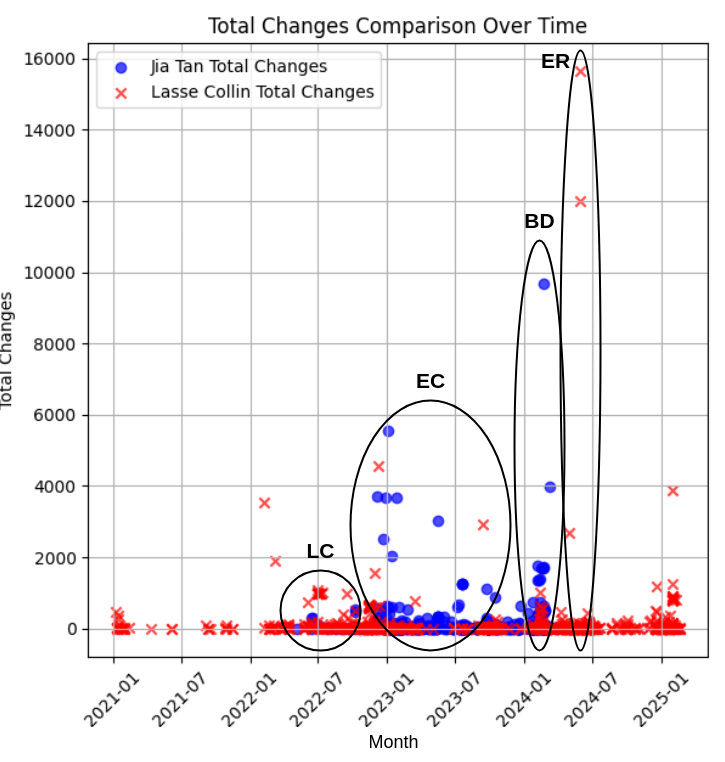}\vspace*{-2.5ex}
    \caption{Total Changes in Commits of authors Jia Tan and Lasse Collin}
    \label{fig:commitChangeScatter}
\end{figure}

\begin{figure}
    \centering\vspace*{-1.75ex}
    \includegraphics[scale=0.34]{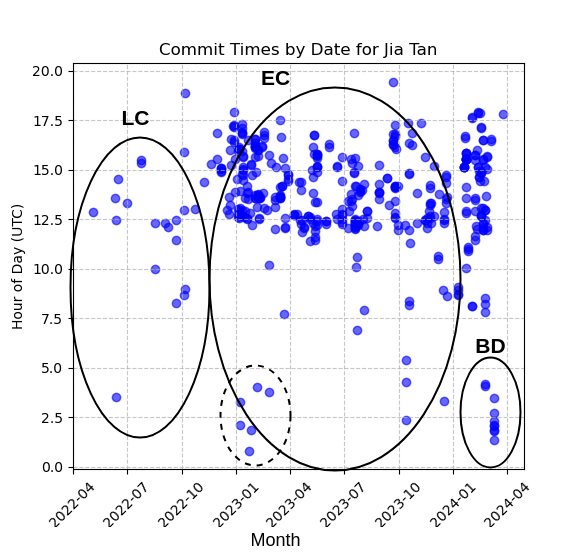}\vspace*{-2.5ex}
    \caption{Author Jia Tan commits plotted over months and time of day.}
    \label{fig:jiatTimeCommits}
\end{figure}

Figure~\ref{fig:commitChangeScatter} plots the average overall changes over time and clusters of the attack stages are highlighted: the most substantial changes occurred when backdoor was inserted [\ac{BD}][\ac{ER}]. The results illustrate that at similar times to the anomalous commit times (see later discussion of \figurename{}~\ref{fig:jiatTimeCommits}), \emph{Jia Tan's} changes were both substantial and anomalous compared to the legitimate maintainer \emph{Lasse Collin}.

\begin{figure}
    \centering
    \includegraphics[scale=0.33]{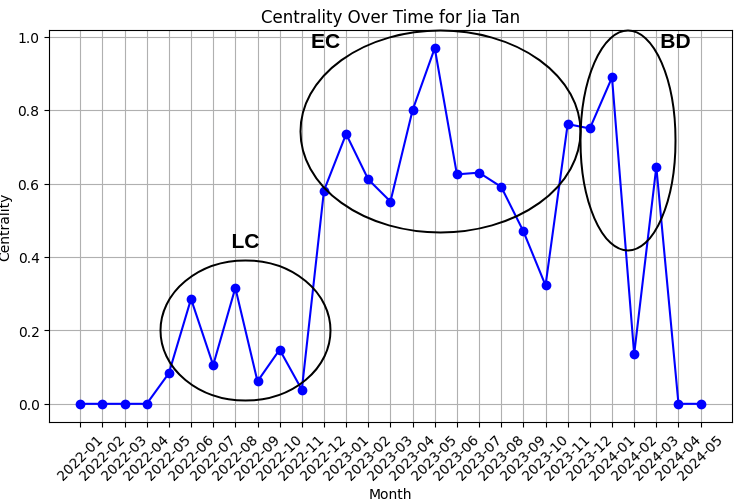}\vspace*{-2ex}
    \caption{Centrality of author Jia Tan}\vspace*{-2.1ex}
    \label{fig:JiaTCentrality}
\end{figure}

To compare \emph{Jia Tan's} impact upon the repository against all other developers, we built a network of the commits to each file and their authors. We then calculate node centrality (indicating the importance of an node (=author) in the network~\cite{borgatti2005centrality}) over time for \emph{Jia Tan}, shown in \figurename{}~\ref{fig:JiaTCentrality}. A higher value of centrality indicates that \emph{Jia Tan's} influence upon the entire source code repository is high, with notable peaks at similar times to the malicious code entries [\ac{EC}][\ac{BD}].

The malicious contributions occurred at atypical times~\cite{Kaspersky2024_2} which might indicate more than one person inserting the code due to working at a different times of day. Figure~\ref{fig:jiatTimeCommits} plots \emph{Jia Tan's} commits over time against the time of day. The cluster in the bottom right corner is the final malicious backdoor insertion [BD]. However, benign code had occasionally been submitted at atypical times of day (e.g.\ \mbox{2023-01}), so alone, these results might not be seen as anomalous.

Further suspicious activities were related to the communications on the mailing list whereby \emph{Lasse Collin} was subjected to a coordinated pressure campaign from multiple sock puppet accounts to relinquish control of the maintenance to \emph{Jia Tan}~[\ac{EC}]. While data points in the mailing lists relating to these authors were limited, an analysis of the pressuring actors' sentiment, affinity for the suspicious actor \emph{Jia Tan} and low interaction with the repository could indicate anomalous and untrustworthy actions. As an example, \emph{Hans Jensen} entered the scene mid-2023 submitting a pair of patches that introduced the \textit{ifunc} feature. These patches were reworked by \emph{Lasse Collin} and then merged by \emph{Jia Tan}. Beyond these interactions, and later filing a Debian bug requesting an update to XZ Utils v5.6.1, \emph{Hans Jensen} does not exist anywhere else on the internet~\cite{Thompson2024, Kaspersky2024_2}. 

\begin{figure}
    \centering\vspace*{-1.5ex}
    \includegraphics[scale=0.36]{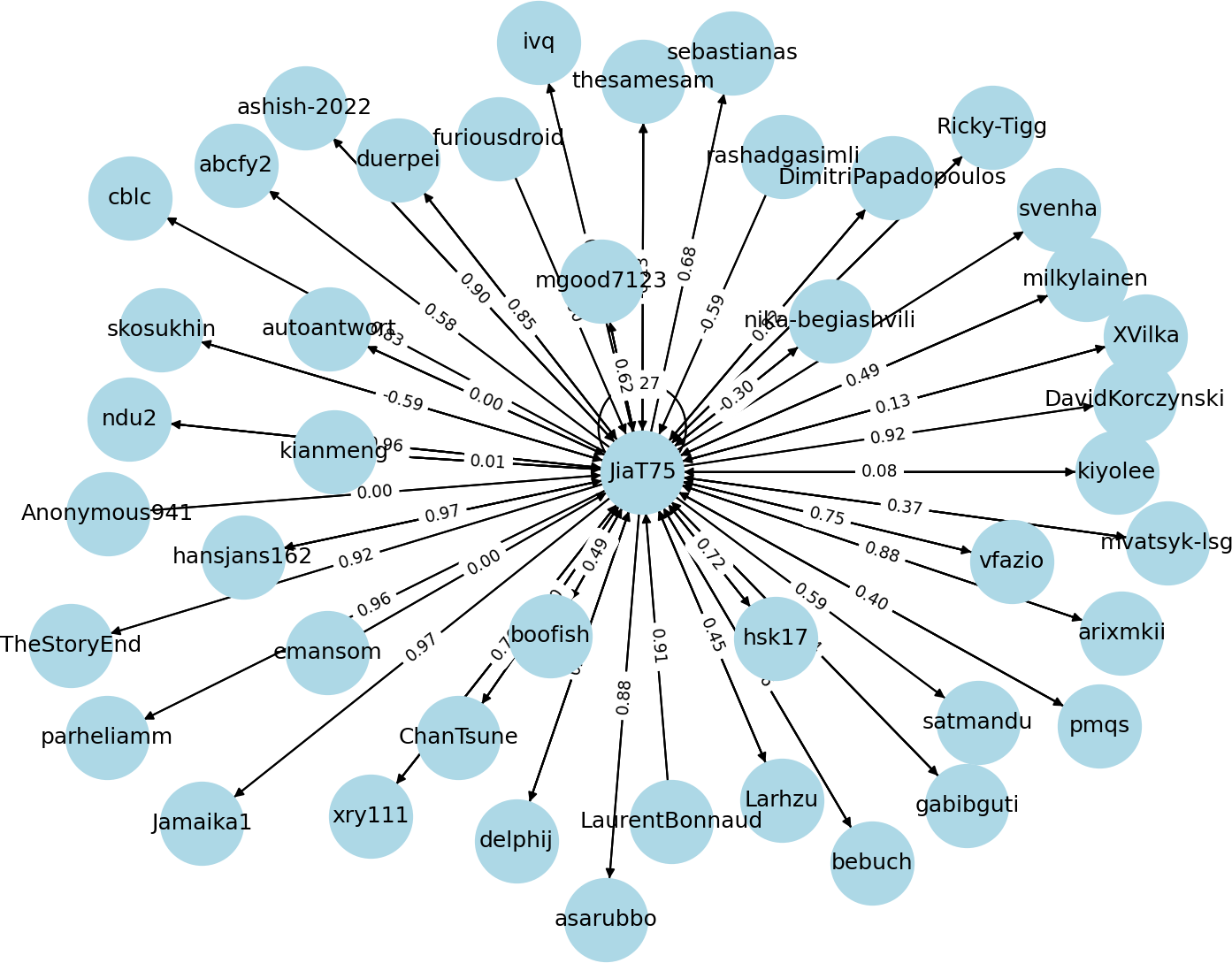}\vspace*{-1.5ex}
    \caption{Mean of the communication sentiment of author Jia Tan.}\vspace*{-2.1ex}
    \label{fig:jiatCentMean}
\end{figure}

Therefore, to analyse this communication, 
\figurename~\ref{fig:jiatCentMean} presents a graph of \emph{Jia Tan's} communication with other developers in the issues created in GitHub. The attribute on each edge is the mean value of the sentiment of that communication. Although this information alone is useful in identifying change in developer interactions, correlating it with change in components is necessary to identify any potential impact.

\begin{figure}
    \centering
    \includegraphics[scale=0.45]{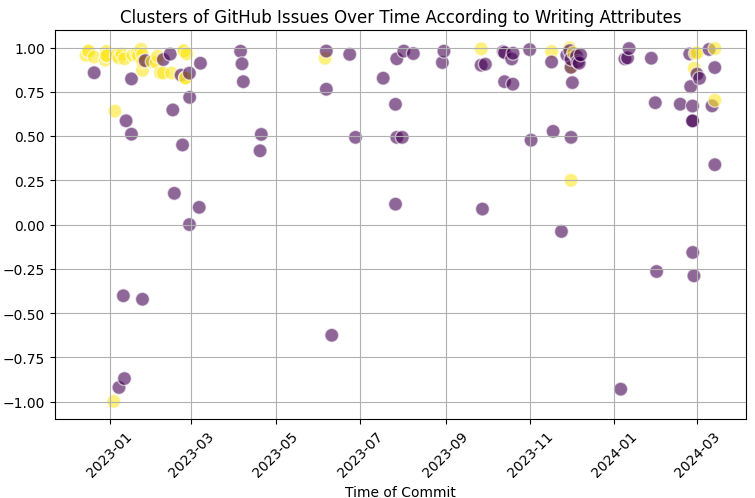}\vspace*{-2ex}
    \caption{Author Jia Tan issues' are clustered (yellow/purple) according to various writing features and plotted against the  sentiment (y-axis) to identify if more than one person was using the account.}\vspace*{-1.85ex}
    \label{fig:jiatIssueClusters}
\end{figure}

To further investigate whether \emph{Jia Tan} was represented by more than one person, we examined various syntactic and lexical features of their writing in the GitHub's issues including: sentiment polarity, subjectivity, lexical richness, punctuation frequency, readability, and average sentence length using the Natural language toolkit \cite{bird2009natural} and TextBlob \cite{loria2018textblob}. 
We then used the \emph{k-means} algorithm (via sci-kit learn \cite{scikit-learn}) to cluster these writing features to identify whether there are several clusters of writing style, indicating that multiple authors where behind the username \emph{Jia Tan}. Experimenting with different $k$ values, i.e.\ the number of clusters, $k=2$ separated clusters best, suggesting that two different persons might have been involved.
Figure~\ref{fig:jiatIssueClusters} plots for these two clusters over time against the sentiment of the issue text (y axis), i.e.\ the multi-dimensional values of the writing features are themselves are not depicted, only the two colours yellow and purple are used to distinguish the two assumed authors. The first cluster (yellow points), potentially indicating one author, has mainly a positive sentiment and is mainly found during escalation of control [\ac{EC}], in particular in early 2023. The second cluster (purple points), potentially another author, takes over the majority of discussions starting from mid 2023. This cluster contains the majority of the negative sentiment while the pressure was being asserted (except for some outliers). The first author returns during the time the final backdoor is inserted in early 2024 [\ac{BD}].

Examining these figures as a whole, it can be see that during the [\ac{EC}] [\ac{BD}] stages, \emph{Jia Tan} has anomalous amount of lines changes in files (\figurename~\ref{fig:commitChangeScatter}), increasing influence on the overall code base (\figurename~\ref{fig:JiaTCentrality}), anomalous times of commits (\figurename~\ref{fig:jiatTimeCommits}),  and potentially multiple authors with increasing negative sentiment in the issues that they created (\figurename~\ref{fig:jiatIssueClusters}). Therefore, 
this analysis shows that integrating technical and social data and analysing the trends might be used to identify anomalous developer behaviour which could indicate subversion and threats. 

\section{Socio-Technical Adaptive Threat Detection}\label{sec:Socio-Technical Adaptive Threat Detection}
\begin{figure}
    \centering\vspace*{-1.5ex}
    \includegraphics[scale=0.6]{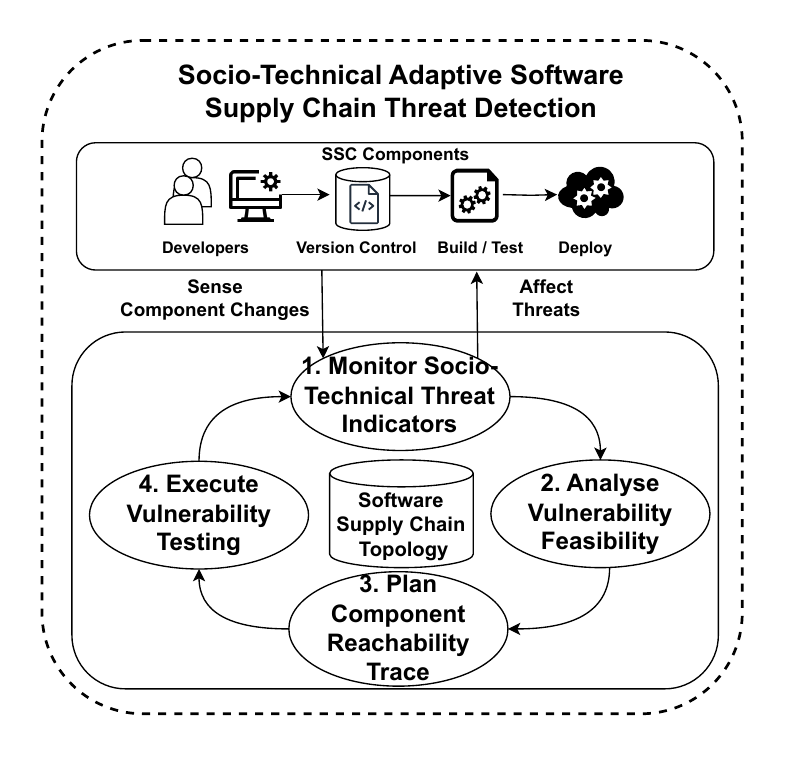}\vspace*{-3.85ex}
    \caption{\ac{SSC} Threat Detection mapped on the \ac{MAPE-K} framework}\vspace*{-1.85ex}
    \label{fig:adaptivessc}
\end{figure}

In this section, we propose \textit{Socio-Technical Adaptive Threat Detection}. 
Figure~\ref{fig:adaptivessc} illustrates the approach by means of the \ac{MAPE-K} adaptive framework \cite{quin2021decentralized}. It will mine data sources (e.g.\ from version control systems, developer chat) to build and maintain socio-technical topologies of the supply chain (e.g.\ from code, tools, developer behaviour). The maintenance of these topologies will allow \textit{monitoring} for change in social and technical components. 
Components where the dynamics between both social and technical components indicate a combined socio-technical vulnerability can then be \textit{analysed} for the level of deviation from a previously known secure state (e.g.\ metrics such as lines of code and code churn). The threat to the \ac{SSC} can then be identified through \textit{planning} further \emph{execution} of testing by tracing these components to determine further insecure states.

\noindent\textbf{Building Socio-Technical Topology Models.} Key to the approach is the ability to manage the structural complexity of the \ac{SSC}. This requires \textit{monitoring} the entire \ac{SSC} for changes to prioritise components for vulnerability analysis. This includes modelling the \textit{connectivity} between components, understanding the interplay between social and technical dimensions, and the \textit{reachability} of vulnerabilities across the different components. To achieve this, we propose the use of multidimensional topologies -- structural models -- of the \ac{SSC} \cite{welsh2022topology}. To build these topologies, a variety of available data sources can be used to model them as graphs. 

Formally, in this approach, a socio-technical topology model of a \ac{SSC} is defined as a tuple  $STT = (S,D,A,R,P,W,M,K,I)$ where:
\begin{itemize}
    \item $S$ is a set of source files S = $\{s_1,s_2,..s_n\}$ 
    \item $D$ a set of tuples representing dependency relationships between source files,  $D  \subseteq S \times S = \{(s_1,s_3),(s_1,s_2)...\}$.
    \item $A$ is a set of authors who contribute to the software repository  $A = \{a_1,a_2...,a_n\}$.
    \item $R$ is a set of tuples defining relationships between authors  $R \subseteq A  \times A = \{(a_1,a_2),(a_1,a_3)\}$.
    \item $P$ as a set of parameters that describe author relationships, such as communication frequency and sentiment, e.g.\  $P = \{frequency,sentiment,...\}$
    \item $W$ is a function assigning a weight to each relationship based on parameters  $W : R \to \mathbb{R}^{|P|}$, e.g.\ $W(a_1,a_2) = (0.2,0.9,...)$ 
    \item $M$ is a relation between  maintainers and source files: $M \subseteq  A \times S  = \{(a_1,s_1),(a_1,s_2),...\}$.
    \item $K$ is a set of activities performed by maintainers while interacting with source files, e.g.\ $K=\{additions,delections,...\}$
    \item $I$ is a function defining the influence of maintainers on source files, based on their activities: $I : M \to \mathbb{N}^{|K|}$, e.g.\ $I(a_1,s_1) = (4,12,...)$.

\end{itemize}

\noindent\textbf{1. Monitoring Socio-Technical Topologies for Threat Indicators.} During this stage, data sources (e.g.\ version control systems, mailing lists) are mined to update the topology models over time. This timeline allows tracking change in the social and technical dimensions. Various metrics will be computed (e.g.\ code churn, network properties, communication frequency and sentiment, volatility of dependencies). Metrics changing together might indicate anomalous behaviour. An actor whose sentiment changes in tandem with changes in coding quality, commit time or overall influence on the code. The difference between $STTs$ at different times can be taken, e.g.\ $\Delta STT = STT(t_2) - STT(t_1)$. To identify suspect components they can be filtered, i.e.\ to select source files according to the developer influence and their changing relationship with other developers
\begin{multline}
S_R = \{s_i \mid \exists (a_1, a_2) \in R_2 - R_1 \text{ such that } s_i 
\\ \text{ is involved in } (a_1, a_2) \text{ and } W(a_1, a_2) \geq W_{\text{threshold}} \}
\end{multline}

and to filter according to their influence on a file:
\begin{multline}
   S_I = \{s_i \mid \exists (a_j, s_i) \in I_2 - I_1 \text{ such that } \\
   I(a_j, s_i) \geq I_{\text{threshold}} \} 
\end{multline}

Then, to select components which fulfil both criteria:  $ S_{\text{selected}} = S_R \cap S_I$. 
$W_\text{threshold}$ and $I_\text{threshold}$ would need to be determined through monitoring the nominal behaviour of the system, as with any anomalous detection system.
 
\noindent\textbf{2. Analysing Components for Vulnerability Feasibility.} The filtered components identified previously will be combined with associated socio-technical threat indicators to analyse the feasibility of one or more vulnerability class existing including: 1) Design flaws such as through architectural model-checking and antipatterns, e.g.\ from Common Weakness Enumerators\footnote{\url{https://cwe.mitre.org/}}. 2) Implementation bugs using static analysis and known vulnerable code, e.g.\ from OWASP TOP~10\footnote{\url{https://owasp.org/www-project-top-ten/}}. \mbox{3) Configuration} errors through validation of insecure default settings. 4) Operation and Maintenance, e.g.\ through dependency auditing and CI/CD logs. 

\noindent\textbf{3. Planning Component Vulnerability Reachability Trace.} If a feasible vulnerability is identified in one of the selected components in the previous stage, the \textit{reachability} of that vulnerability must then be determined through tracing its impact upon downstream and upstream components in the SSC. The first stage can simply identify adjacent components where $D^{+}_{s}$ is the set of adjacent upstream components in the form $\{p',p\}$ and $D^{-}_{s}$ the set of downstream components of the form $\{p,p'\}$. 

\noindent\textbf{4. Executing Vulnerability Testing.} An analysis will need to test identified components, pruning elements of the dependability tree where the feasibility is low. This testing could take several forms, such as analysing the syntax and semantics of the code, e.g.\ through static or dynamic analysis, fuzz testing or machine learning approaches where the particular approach will need to be evaluated and selected according to features such as data availability.


\section{Limitations and Research Vision}\label{sec:Limitations and Research Vision}
In this section, we present a discussion of the proposed approach and identify future research challenges to achieve it.
\textbf{Adaptation.} We argue that threat detection in \acp{SSC} must be adaptive to manage its high rate of change. While adaptative software frameworks are proven in many domains, \acp{SSC} pose challenges due to the heterogeneity and distributed nature of the components. The adaptation rate must consider this rate of change to be effective and the control functionality will need to accommodate the decentralised supply chain structure. Where engineering decentralised adaptation is an on-going research challenge \cite{quin2021decentralized}. \\
\noindent\textbf{Developer Behaviour Analysis.}
Developer activities as indicators for vulnerabilities have been investigated previously (Section~\ref{sec:Background and Related Work}). However, these sources are large, varied and continuously evolving, e.g.\ new tools such as generative AI. To achieve the vision of this paper, we need to understand: \textit{What types of developer behaviour can support identification of insecure states in \ac{SSC} components? and How does the efficacy of these approaches vary between \ac{SSC} components?} Eliciting the appropriate technique for each software repository will require context-aware methods.\\ 
\noindent\textbf{Software Component Analysis.}
Analysing individual software components for vulnerabilities has seen considerable work in literature (Section~\ref{sec:Background and Related Work}), although less so considering interdependency between components. New modelling approaches which apply metrics for vulnerability feasibility, suitable for comparison across diverse components are needed. Hence, to achieve this we seek to understand: \textit{What are the patterns which can be used to identify sequences of insecure states in \ac{SSC} components?} Identifying patterns of connected, potentially vulnerable components may support efficient identification of vulnerabilities without exhaustive search. \\
\textbf{Evaluation of the Approach.} Identifying \ac{SSC} attacks is challenging due to sparse attack data, although this is increasing. In part, we mitigate this by identifying suspicious activities used to inform targeted vulnerability analysis. Regardless, we seek to understand: \textit{How effectively does an adaptive approach support socio-technical threat detection in \acp{SSC}?} Comparing adaptive and non-adaptive approaches will require a controlled test bed covering the complete supply chain from design to deployment. The creation of such a test bed will strongly benefit \ac{SSC} security research. Yet, while some open-source data is available, diverse social data and tool configurations will need to be developed.

\section{Conclusion}\label{sec:Conclusion}
In this position paper, we outlined our research vision of a \textit{Socio-Technical Adaptive SSC Threat Detection}. We motivated the need to build socio-technical topologies to understand the relationship between socio-technical dynamics and vulnerabilities through analysis of the data from the XZ Utils attack. We conclude with several challenges to the realisation of this approach, including investigating social and technical analysis techniques and the need for a \ac{SSC} testbed to support research and evaluation efforts.

\section*{Acknowledgements}
This project has received co-funding from the European Union's Digital Europe Programme under grant agreement no.\ 101127453 National Coordination Centre for Cybersecurity in Iceland and 101127307 Defend Iceland: Nationwide bug bounty platform. 

\bibliographystyle{IEEEtran}
\bibliography{refs}

\end{document}